\documentclass[onecollarge,natbib]{svjour2}
\bibpunct{[}{]}{;}{n}{}{,} 
\smartqed  
\usepackage{graphicx}
\journalname{Few-Body Systems (EFB22)}
\begin{document}

\title{$\phi N$ photoproduction coupled with the $K\Lambda^*$
  channel~\thanks{The present work is supported by Basic Science
    Research Program through the National Research Foundation of Korea
    funded by the Ministry of Education, Science and 
Technology (Grant Number:  2012001083).  It is also supported in part
by the Grant-in-Aid for Scientific Research on Priority Areas titled
``Elucidation of New Hadrons with a Variety of
Flavors''(E01:21105006).} 
}


\author{H.-Ch. Kim         \and H.-Y. Ryu \and A. Titov \and A. Hosaka
}


\institute{H.-Ch. Kim \at
              Department of Physics, Inha University, Incheon
              402--751, Republic of Korea \\
              Tel.: +82-32-860-7652\\
              Fax:  +82-32-872-7561\\
              \email{hchkim@inha.ac.kr}           
           \and
           H.-Y. Ryu \at
              Korea Institute of Science and Technology Information
              (KISTI) 245, Daejeon 305-806, Korea\\
              \email{hyryu@rcnp.osaka-u.ac.jp}
           \and
           A. Titov \at
           Bogoliubov Laboratory of Theoretical Physics, JINR, Dubna
           141980, Russia\\
           \email{atitov@theor.jinr.ru}
           \and
           A. Hosaka \at
           Research Center for Nuclear Physics, Osaka University,
  Ibaraki 567--0047, Japan\\
\email{hosaka@rcnp.osaka-u.ac.jp}
}

\date{Received: date / Accepted: date}

\maketitle

\begin{abstract}
We present in this talk a recent investigation on $\phi$
photoproduction, emphasizing the rescattering effects of the
$K\Lambda^*$ channel near the threshold region. We discuss the results
of the differential cross section and the angular distributions.  
\end{abstract}

\section{Introduction}
\label{intro}
The strangeness content of the nucleon has been one of the most
interesting issues. In particular, it is of great importance to
understand how strange quarks can be created from the
\textit{nonstrange} nucleon. In this sense, $\phi$ photoproduction 
provides essential information on it. On the other hand, the coupling
of the $\phi$ meson to the nucleon is suppressed by the
Okubo-Zweig-Iizuka (OZI) rule. Because of this, the pomeron was
regarded as the main contribution to $\phi$ photoproduction.
However, a recent measurement performed by the LEPS
collaboration~\cite{Mibe:2005er} finds a bump-like 
structure around the photon energy $E_\gamma \approx 2.3$ GeV. It
seems that the Pomeron alone cannot account for this bump-like
structure and requires that one should consider other production
mechanism of $\phi$ photoproduction near the threshold energy.  

In an attempt to explain this experiment, Ozaki et
al.~\cite{ozaki2009} introduced coupled-channel effects with the
$K$-matrix method. They considered the $\gamma N\to K\Lambda^*(1520)$   
and $K\Lambda^*\to\phi N$ kernels~\cite{Nam:2005uq} but the
coupled-channel effects turned out to be not enough to describe 
the bump-like structure at $E_\gamma \approx 2.3$ GeV. Another
idea was suggested by Ref.~\cite{Kiswandhi:2010ub}: the
bump-like structure could arise from a destructive interference
due to the $N^*(2010)$ resonance with the large $s\bar{s}$ content.      

In the present talk, we will briefly demonstrate a recent
investigation~\cite{Ryu:2012tw} in which it was shown that the
$K\Lambda^*$ rescattering effects may explain the $2.3$-GeV bump-like
structure on the contrary to Ref.~\cite{ozaki2009}, emphasizing in
particular the rescattering effects of the $K\Lambda^*$ channel.  

\section{Effects of the  $K^+  \Lambda(1520)$ channel}
\label{sec:2}
In Ref.~\cite{Ryu:2012tw}, the amplitude of $\phi$ photoproduction was
constructed by considering the Pomeron-exchange, pseudoscalar
meson-exchanges, and seven different rescattering amplitudes. It was
found that the $K\Lambda(1520)$ rescattering amplitude is the dominant
one among others in describing $\phi$ photoproduction near the
threshold. Thus, we will focus on the $K^+\Lambda(1520)$ channel in
this talk. The main component of the $K^+ \Lambda(1520)$ rescattering
effects is surely the amplitude of the $K^+\Lambda^*$ 
photoproduction~\cite{Nam:2005uq}. Then, the rescattering equation can
be written as 
\begin{eqnarray}
\mathcal{M}_{\gamma N\to \phi N} (p,p';s) &=&
\mathcal{M}^{\mathrm{Born}}_{\gamma N\to \phi N} 
(p,p';s) \cr &&\hspace{-2cm} +\; \int d^3 q
\frac{\omega + E}{(2\pi)^3 2\omega  E} \mathcal{M}_{\gamma N\to K^+
  \Lambda^*} (p,q;s) \frac1{s-(\omega+E)^2 + i\varepsilon} 
\mathcal{M}_{K^+\Lambda^* \to \phi N} (q,p';s),    
\label{eq:rescat}
\end{eqnarray}
which is a Blankenbecler-Sugar (BbS)
equation~\cite{Blankenbecler:1965gx}.  The amplitude
$\mathcal{M}^{\mathrm{Born}}_{\gamma N\to \phi N}$ contains the
diagrams of pomeron- and meson-exchanges. The second part of
Eq.(\ref{eq:rescat}) in the right-hand side denotes the rescattering
amplitude of the $K\Lambda^*$ channel. Both $\mathcal{M}_{\gamma
  N\to K^+\Lambda^*} (p,k;s)$ and $\mathcal{M}_{K^+\Lambda^* \to \phi
  N} (k,p';s)$ are the off-mass-shell extended amplitudes for the
$\gamma p \to K^+\Lambda^*$ and $K^+\Lambda^*\to \phi p$,
respectively. $\omega$ and $E$ correspond to the off-mass-shell energies
of the $K^+$ and the $\Lambda^*$ .in the intermediate states.
$s$ is a Mandelstam variable, i.e. the square of the total energy
$s=(E_\gamma +E_p)^2$.     

Since it is quite complicated to deal with Eq.(\ref{eq:rescat}) in a
full coupled-channel formalism, we will first concentrate on the
imaginary part of Eq.(\ref{eq:rescat}), which can be easily derived by
the two-body unitarity relation (Landau-Cutkosky rule). In this case,
we need only the on-mass-shell amplitudes. The full calculation of
Eq.(\ref{eq:rescat}) is under investigation and will be presented
elsewhere. The imaginary part of the BbS equation is written as     
\begin{equation}
  \label{eq:10}
\mathrm{Im}\mathcal{ M}_{K^+ \Lambda^* \mathrm{rescatt.}}
\;=\; -\frac{1}{8\pi}\frac{r}{\sqrt{s}} \int \frac{d \Omega}{4 \pi}
\mathcal{ M}_L(\gamma p \to K^+ \Lambda^*) \mathcal{ M}_R^\dagger (K^+ 
\Lambda^* \to \phi p),
\end{equation}
where $r$ is the magnitude of the $K^+$ on-mass-shell three
momentum. For detailed formailsm, we refer to the recent
work~\cite{Ryu:2012tw}. 

\section{Results}
\begin{figure}[h]
\centering
       \includegraphics[scale=0.5]{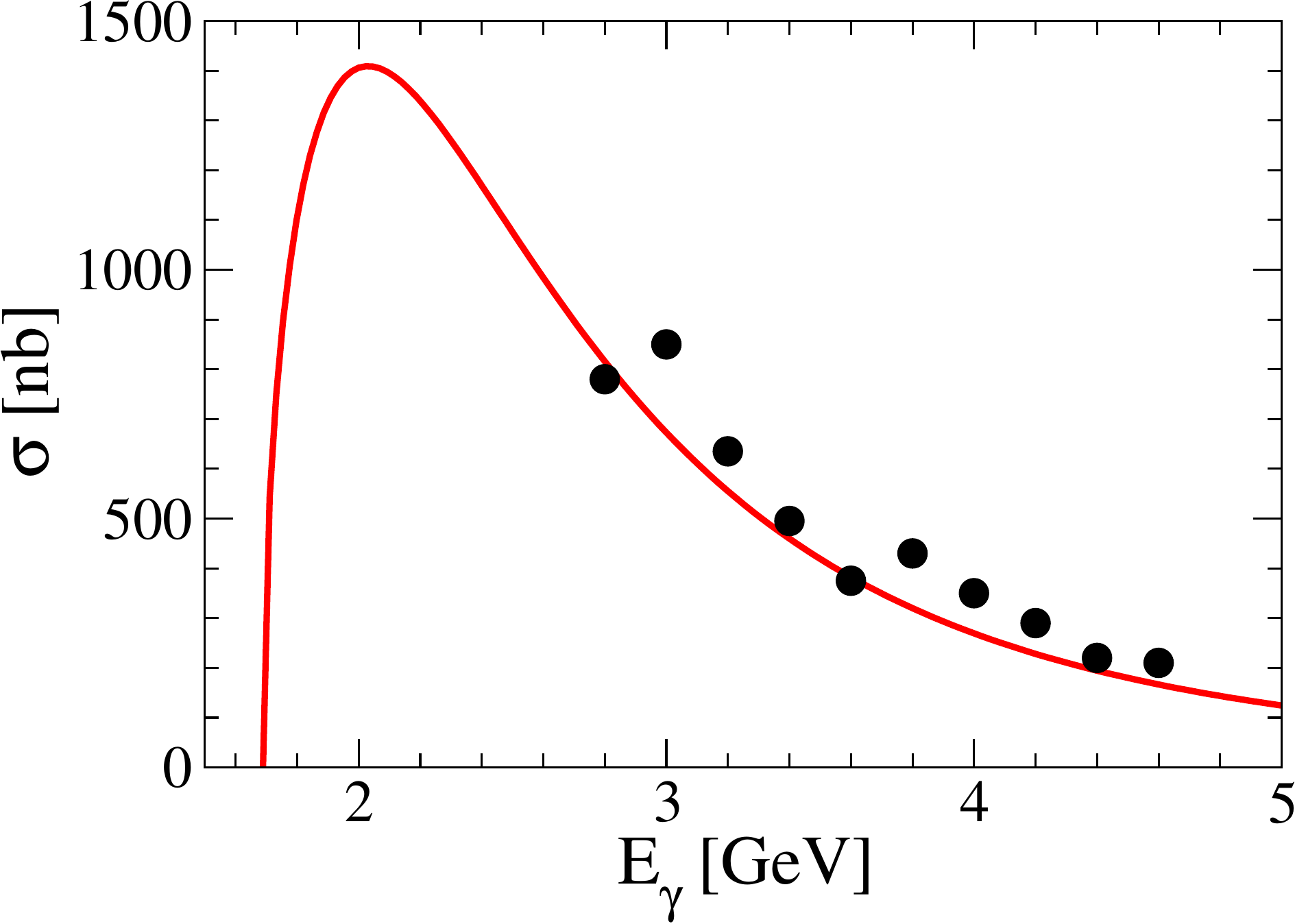}
    \caption{Total cross section for the $\gamma p \to K^+
      \Lambda(1520)$ process. The experimental data taken from
      Ref.~\cite{Adelseck:1986fb}.} 
  \label{fig:1}      
\end{figure}
We now present the results for the rescattering effects of the
$K\Lambda^*$ coupled channel on $\phi$ photoproduction. 
Figure~\ref{fig:1} draws the results of the total cross section for
$K^+ \Lambda(1520)$ photoproduction, based on Ref.~\cite{Nam:2005uq}. 
The experimental data are well repdoduced. Employing the $\gamma p \to
K^+ \Lambda(1520)$ amplitude, we are able to consider the $K\Lambda^*$
rescattering in $\phi$ photoproduction.  

\begin{figure}[ht]
\centering
       \includegraphics[scale=0.6]{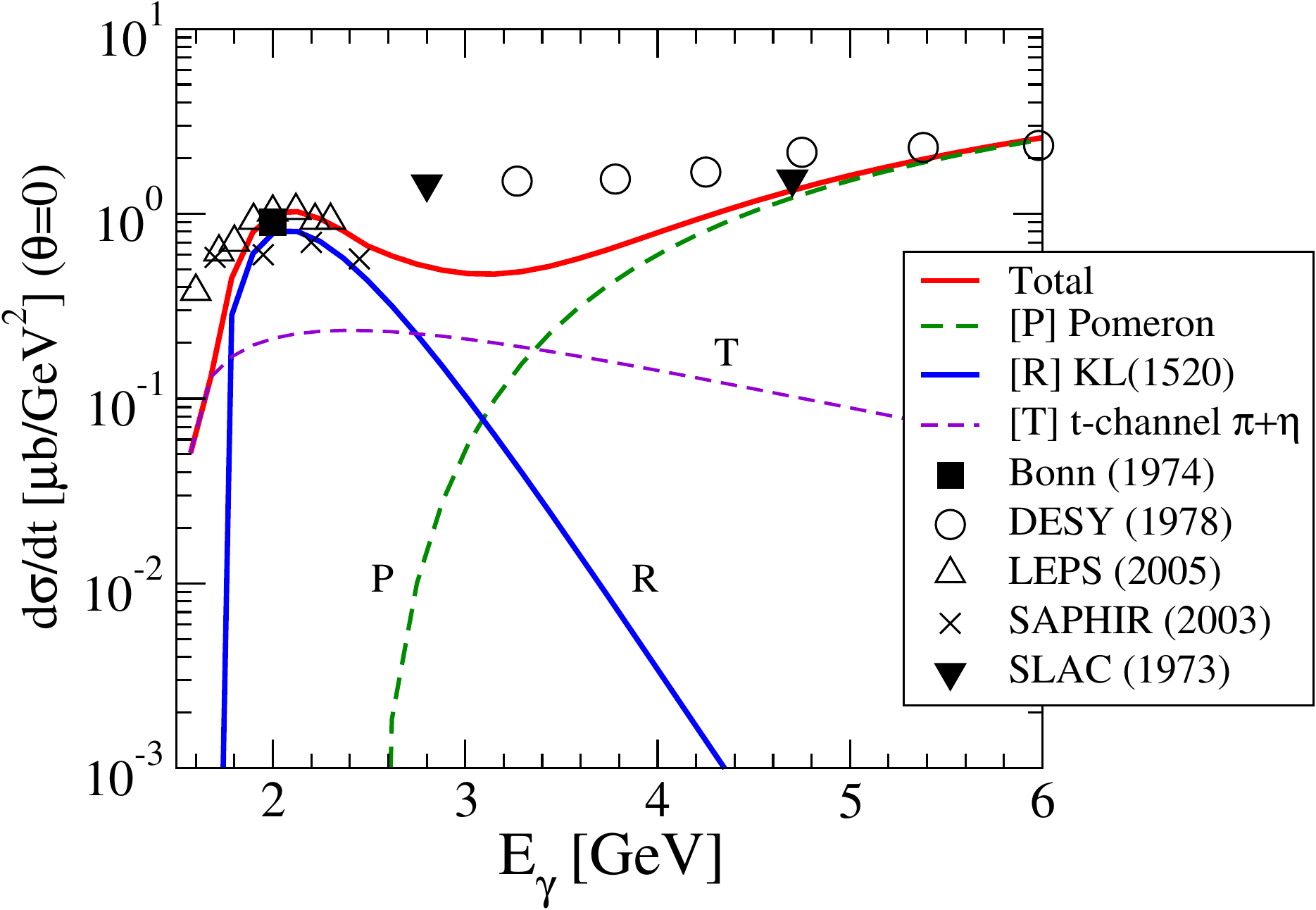}
    \caption{(Color on-line) Differential cross section as a function
      of the photon energy $E_{\gamma}$ in a log scale. The thick
      solid curve depicts the result with all contributions
      included. The solid curves with the symbols $P$, $R$, and $T$
      denote the Pomeron contribution, $K\Lambda^*$ rescattering
      effects, and the $t$-channel contribution of $\pi$- and
      $\eta$-exchanges. }  
  \label{fig:2}      
\end{figure}

In Fig.~\ref{fig:2}, we show each contribution to the differential
cross section $d\sigma/dt$ for $\phi$ photoproduction in log
scale. The dashed curve with P represents the pomeron-exchange 
contribution. The pomeron governs typically the general $E_\gamma$ 
dependence, in particular, in the high energy region, while the
$t$-channel effects designated by T contribute to the differential
cross section almost equally along $E_\gamma$. The $K\Lambda(1520)$
rescattering effects start to arise from the threshold drastically
till around 2 GeV, then fall off fast. Thus, the interference of all
these three contributions make it possible to describe the bum-like
structure around $E_\gamma =2.3$ GeV. 

\begin{figure}[h]
\centering
       \includegraphics[scale=0.7, angle=-90]{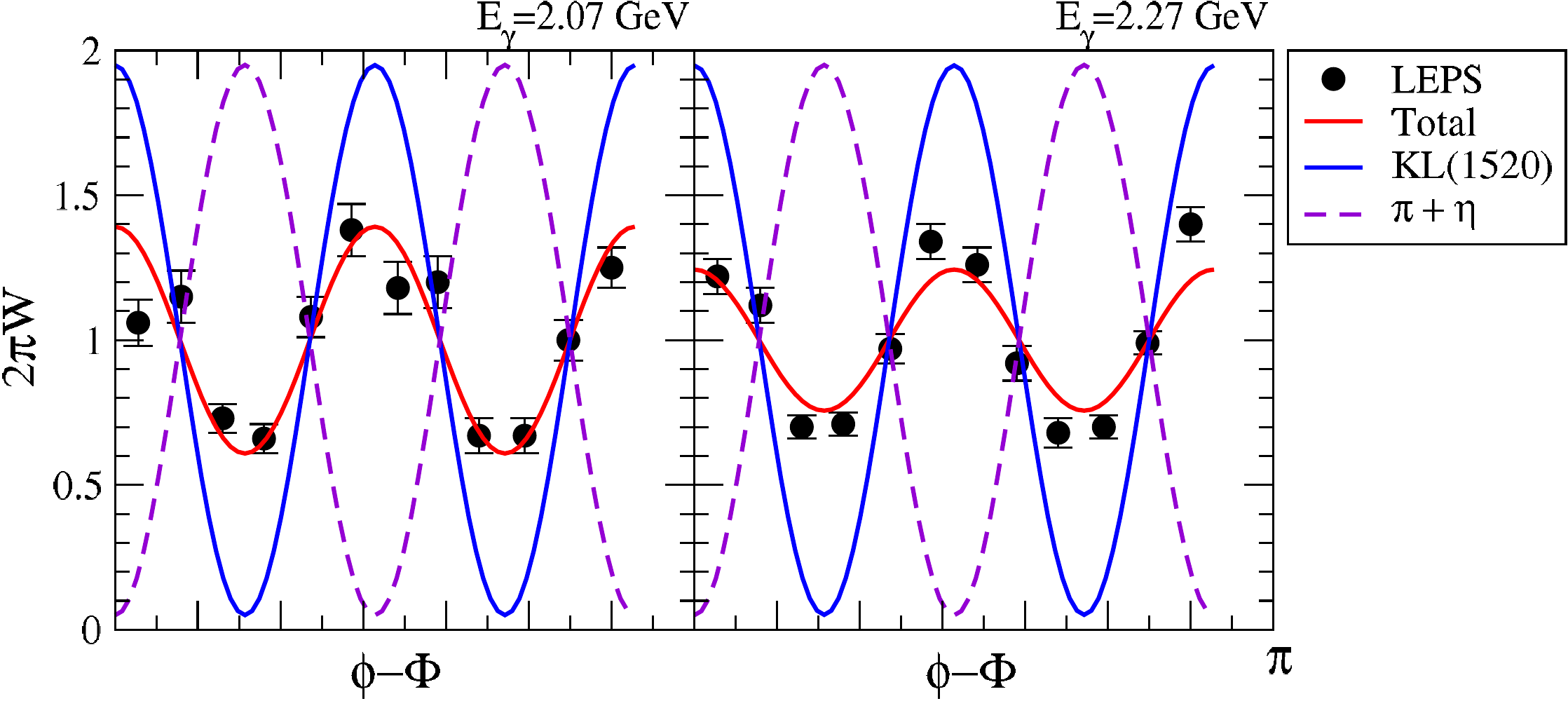}
    \caption{(Color on-line) The decay angular distributions for
$-0.2<t+|t|_{\mathrm{min}}$. The experimental data are taken from
Ref.~\cite{Mibe:2005er}.}   
  \label{fig:3}      
\end{figure}

The angular distributions of the $\phi\to K^+ K^-$ decay in the $\phi$
rest frame allow one to get access to the helicity amplitudes 
experimentally~\cite{Gottfried:1964nx,Schilling:1969um}.
They were measured by the LEPS collaboration at 
forward angles ($-0.2 <t+|t|_{\mathrm{min}}$) in two different energy
regions: $1.97<E_\gamma< 2.17$ GeV and $2.17 <E_\gamma<2.37$  
GeV~\cite{Mibe:2005er}. Here, $|t|_{\rm min}$ denotes the minimum
four-momentum transfer from the incident photon to the $\phi$
meson. In this talk, we focus on the one-dimensional decay angular
distribution $2\pi W(\phi-\Phi)$, since it illuminates 
the effects of the $K\Lambda^*$ coupled channel. The $2\pi W$ is
defined as   
\begin{equation}
  \label{eq:3}
 2\pi W(\phi -\Phi) \;=\;  1 + 2 P_\gamma  \overline{\rho}^{1}_{1-1}
\cos 2(\phi -\Phi), 
\end{equation}
where $\phi$ is the polar and azimuthal angles of the decay particle
$K^+$ in the $\phi$ rest frame. $\Phi$ stands for the azimuthal angle
of the photon polarization in the center-of-mass frame. $P_\gamma$
denotes the degree of the polarization of the photon beam. The
definition of $\overline{\rho}^{1}_{1-1}$ can be found in
Ref.~\cite{Ryu:2012tw}. In Fig.~\ref{fig:3} we compare the results of
the $2\pi W(\phi-\Phi)$ with the experimental data. Note that since
the photon energy $E_\gamma=2.07$ GeV is small, the pomeron does not
come into play. Interestingly, the $K\Lambda^*$ rescattering effects
turn out to be crucial in describing $2\pi W(\phi-\Phi)$. The
$t$-channel contribution interferes destructively with the
$K\Lambda^*$ effects.  

\section{Summary and outlook}
In the present talk, we briefly reviewed a recent investigation on
the $K\Lambda^*$ coupled-channel effects in addition to the
conventional approach of Pomeron-, $\pi$-, and $\eta$-exchanges. 
We found that the $K\Lambda(1520)$ rescattering effects play a crucial
role in describing  the bump-like structure near $E_\gamma\approx 2.3$
GeV of the LEPS experiment. The angular distribution of the
$\phi$ also was well explained by the inclusion of the
$K\Lambda(1520)$ coupled channel.  

In this work, we have considered only the imaginary part of the
$K\Lambda(1520)$ rescattering effects. However, the real part will be
as equally important as the imaginary one. The corresponding
investigation is under way and will soon appear elsewhere.


\end{document}